\documentstyle[sprocl,psfig,citesort]{article}

\begin{document}

\title{RANDOMNESS AND APPARENT FRACTALITY}

\author{ D.A. LIDAR (HAMBURGER) }
\address{Racah Institute of Physics and Fritz Haber Center for Molecular
Dynamics, \\
the Hebrew University, Jerusalem 91904, Israel, \\
http://www.fh.huji.ac.il/$\sim$dani}

\author{ O. MALCAI,$^{(1)}$ O. BIHAM$^{(2)}$ }
\address{Racah Institute of Physics, the Hebrew University, Jerusalem 91904,
Israel, \\
(1) malcai@flounder.fiz.huji.ac.il, \\
(2) http://www.fiz.huji.ac.il/staff/acc/faculty/biham}

\author{ D. AVNIR }
\address{Institute of Chemistry and Fritz Haber Center for Molecular
Dynamics, \\
the Hebrew University, Jerusalem 91904, Israel,
http://chem.ch.huji.ac.il/employee/avnir/iavnir.htm}

\address{Published in: {\it Proc. Int. Conf. Fractals and Chaos in Chem. Engin.}, M. Giona, ed., Sept. 1996, Roma}

\maketitle

\abstracts{
We show that when the standard techniques for calculating fractal dimensions
in empirical data (such as the box counting) are applied on uniformly random
structures, apparent fractal behavior is observed in a range between
physically relevant cutoffs. This range, spanning between one and two decades
for densities of 0.1 and lower, is in good agreement with the typical range
observed in experiments. The dimensions are not universal and depend on
density. Our observations are applicable to spatial, temporal and spectral
random structures, all with non-zero measure. Fat fractal analysis does not
seem to add information over routine fractal analysis procedures. Most
significantly, we find that this apparent fractal behavior is robust even to
the presence of moderate correlations. We thus propose that apparent fractal
behavior observed experimentally over a limited range in some systems, may
often have its origin in underlying randomness.
}

\section{Introduction}

\begin{figure}
\psfig{figure=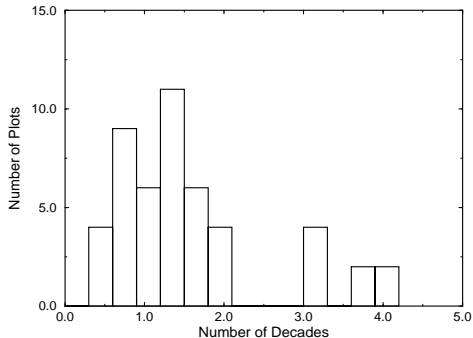,width=12cm,angle=270}
\vskip -1.8cm
\caption{
To obtain a general idea about the experimental status of fractal dimension
measurements we collected all such measurements presented in Ref.[2] and
measured the width of the linear range in the log-log plots (measured in
decades) over which the FD was determined.  This histogram shows the number of
plots as a function of the number of decades of the linear range.  One can see
that most experimental measurements of fractal dimensions are based on data
that extends between one and two decades.  Note further that all the data with
three and four decades, which come from a single paper, is the determination
of the Hurst exponent for temporal and not structural data.\hfill
\hspace{1.0cm}}
\label{fig:fewdecades}
\end{figure}

Fractal structures have been observed in a large variety of experimental
systems in physics, chemistry and
biology.\cite{Mandelbrot,Feder,Avnir:book,Bunde,Stanley,Schroeder} Unlike
exact (mathematical) fractals which are constructed to maintain scale
invariance over many orders of magnitude, and most existing physical models
displaying fractal behavior,\cite{many-decades} for {\em empirical} fractals
the range over which they obey a scaling law is necessarily restricted by
upper and lower cutoffs. In most experimental situations this range may be
quite small, namely not more than one or two orders of magnitude
(Fig.\ref{fig:fewdecades}). Nevertheless, even in these cases the fractal
analysis condenses data into useful relations between different quantities and
often provides valuable insight.

Motivated by the yet inexplicable abundance of reported fractals, we consider
here the apparent fractal properties of systems which are governed by
uniformly random distributions.  The reasons for this choice are several.
First, randomness is abundant in nature. Second, although a uniformly random
system cannot be fully scale invariant, it may, as we show below, display
apparent fractality over a limited range, perhaps in better agreement with the
actual ranges observed than a model which is inherently scale free.  Third, a
model of uniform randomness is a convenient limit, on top of which
correlations can be introduced as perturbations.

\section{The Basic Model}
\label{model}

To illustrate our ideas we use a model that consists of a random distribution
of spheres of diameter $d$, in the limit of low volume fraction occupied by
the spheres.  The positions of the centers of these spheres are determined by
a uniform random distribution and the spheres are allowed to overlap.  This
model may approximately describe the spatial distribution of objects such as
pores in porous media, craters on the moon, droplets in a cloud and adsorbates
on a substrate as well as some energy spectra and random temporal
signals.

\begin{figure}
\psfig{figure=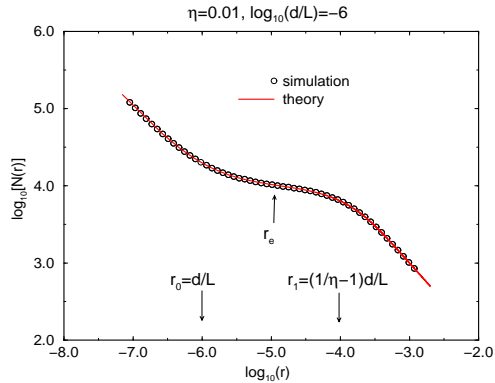,width=12cm,angle=270}
\vskip -1.8cm
\caption{
Comparison of simulation results (circles) to the theoretical prediction of
Eq.(\protect\ref{eq:<N>}) (solid line) for the number of intersected boxes as
a function of their size, for one dimensional penetrable rods. The coverage is
$\eta=0.01$ and the rod length is $d/L=10^\protect{-6\protect}$. The cutoffs
are manifested as the two knees in the graph. The lower bound $r_0$ is seen to
be located at $r=d/L$. The upper bound $r_1$ is at $r=(1/\eta-1)d/L$, also
conforming with the prediction in the text. Also indicated is the estimated
middle point $r_e$.\hfill \hspace{1.0cm}}
\label{fig:Nresults}
\end{figure}

To simplify the analysis we consider here (without loss of generality) the one
dimensional case, where the spheres are $M$ rods of length $d$ which are
placed on a line of length $L \gg d$. The positions of the rod centers are
determined by a uniform random distribution. The rods are allowed to overlap
and are positioned with no correlations. An information-theory argument can be
used to show that this distribution is generic, or ``minimal'', in the sense
that it is characteristic of physical processes in which only the first moment
(such as the density) is determined from
outside.\cite{me:random-model,D-comment5} Below we calculate the fractal
dimension (FD) of the resulting set using the {\em box-counting} (BC)
procedure, which is a common techniques for determination of FD in empirical
data.\cite{higherorder} In the BC technique one divides the embedding space
into boxes of linear size $l$. It is convenient to work with the dimensionless
quantity $r \equiv l/L$ for the box size. The number of boxes that have
intersection with the measured object, $N(r)$, is then plotted vs. $r$ on a
log-log scale. The range of $r$ is limited from below by the finest feature in
the object and from above by the entire object size. Apparent fractal behavior
is commonly declared in a range bound between physical cutoffs if the log-log
plot of $N(r)$ vs. $r$ is linear over one or more
decades\cite{Pfeifer-in-Avnir:book} in that range.  The dimension is given by:

\begin{equation}
D = - {\rm slope}\:\{\log(r), \log[N(r)] \}.
\label{eq:D}
\end{equation}

\noindent We will now show that our model generates approximate linearity over
a range which would conventionally be accepted to indicate fractality. The
lower cutoff is given by the rod length,

\begin{equation}
r_0 = d/L ,
\label{eq:r0}
\end{equation}

\noindent since below this scale no new information is obtained by decreasing
the box size. The upper cutoff is determined by the average distance
between adjacent rod edges,

\begin{equation}
r_1 = 1/M-d/L ,
\label{eq:r1}
\end{equation}

\noindent because above this scale (on average) all boxes are occupied. This
allows us to define an {\em estimated} scaling range as:\cite{D-comment3}

\begin{equation}
\Delta_e = \log_{10}(r_1)-\log_{10}(r_0).
\label{eq:Delta}
\end{equation}

\noindent Since the actual value of $N(r)$ depends on the particular set of
random numbers drawn, one can only obtain the expectation value $\langle
N(r)\rangle$.  However, the law of large numbers ensures that for a large
enough number of rods, the deviations from this value will be insignificant.

Following probabilistic arguments of the type used by
Weissberg\cite{Weissberg} and Torquato and Stell,\cite{Torquato:3} one obtains
that out of the total of $1/r$ boxes the number of boxes that intersect the set
is:\cite{sketch}

\begin{equation}
\langle N(r)\rangle = {1\over r} \left\{ 1-[1-(r+d/L)]^{M} \right\}.
\label{eq:<N>}
\end{equation}

\begin{figure}
\psfig{figure=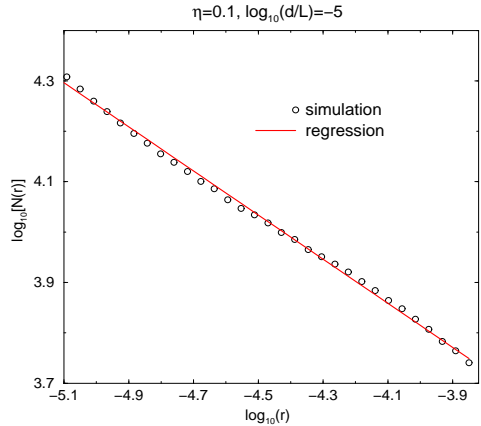,width=12cm,angle=270}
\vskip -1.6cm
\caption{
Simulation results (circles) for the number of intersected boxes $N(r)$
vs. $r$ in the experimentally relevant range (between cutoffs), along with a
linear regression fit for coverage $\eta=0.1$ ($d/L=10^{-5},\,
M=10^4$).\hfill \hspace{1.0cm}}
\label{fig:Nlinear}
\end{figure}

\noindent Simulation results in terms of the {\em coverage} $\eta \equiv M d/L$
are shown in Fig.\ref{fig:Nresults}, along with the theoretical prediction of
Eq.(\ref{eq:<N>}). An excellent agreement is evident.\cite{goodfit}

\begin{figure}
\psfig{figure=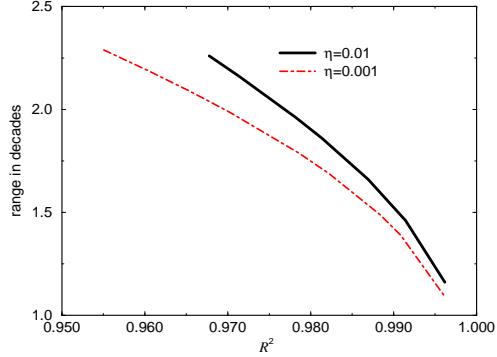,width=12cm,angle=270}
\vskip -1.8cm
\caption{
The range of linearity, $\Delta$, as a function of imposed coefficient of
determination, $R^2$, in a linear regression analysis.\hfill \hspace{1.0cm}}
\label{fig:range}
\end{figure}

Next, we examine the apparent FD [Eq.(\ref{eq:D})], {\it by mimicing the
standard experimental procedure} of using linear regression analysis between
the cutoffs.  The simulation results and the linear fit for $\eta = 0.1$ are
shown in Fig.\ref{fig:Nlinear} for the range which is used to determine
empirical FDs. More than a decade of linearity is observed for this high
coverage. The slight inflexion of the simulation results may be smeared out by
noise in a real experiment. We next evaluate the slopes and {\em actual}
ranges of linearity $\Delta$ (generally $\neq \Delta_e$), under varying
degrees of strictness of linearity, as measured by the coefficient of
determination $R^2$. Typical results are shown in Fig.\ref{fig:range}, where,
e.g. for $\eta=0.01$, more than two decades of linear behavior are exhibited
for a required value of $R^2$ of below 0.975.  This is well within the
experimental norm as most experimental measurements of fractal objects do not
extend for more than two orders of magnitude
(Fig.\ref{fig:fewdecades}). Moreover, this agreement with experimental data is
in contrast to that of most other physical models of fractality, which predict
much larger ranges.\cite{many-decades} Increasing $\eta$ beyond 0.1 results in
a decline of both $\Delta$ and $\Delta_e$ to below one decade and hence the
apparent fractality is {\em restricted to $\eta\leq 0.1$}.

The results of the regression analysis for the apparent FD as a function of
$\eta$ are shown in Fig.\ref{fig:D} and are further compared to an analytical
expression, obtained by calculating the logarithmic derivative of $N(r)$ at
the {\em estimated} middle point $r_e = \sqrt{r_0 r_1}$, in the $M
\rightarrow \infty$, constant coverage limit:\cite{me:random-model}

\begin{equation}
D = 1-{\sqrt{\eta(1-\eta)} \over {\exp \left( \eta+\sqrt{\eta(1-\eta)}
\right)-1}}.
\label{eq:Dresult}
\end{equation}

\noindent As seen in Fig.\ref{fig:D}, the FD predicted by Eq.(\ref{eq:Dresult})
is somewhat lower than the regression result and can serve as a lower bound.
In the limit of small $\eta$, one can further simplify Eq.(\ref{eq:Dresult})
and obtain

\begin{equation}
D \approx \left(\eta \over {1-\eta}\right)^{1/2}, \ \ \ \ \ \ \eta \ll 1.
\end{equation}

\section{Impenetrable Rods}

\begin{figure}
\psfig{figure=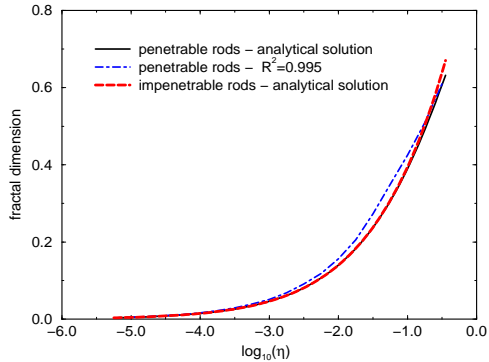,width=12cm,angle=270}
\vskip -1.8cm
\caption{
Apparent fractality (FD) as computed by linear regression with $R^2 = 0.995$
(upper curve). The predictions of the analytical equations,
Eqs.(\protect\ref{eq:Dresult}) and (\protect\ref{eq:Dhard}) (for the
penetrable and impenetrable rods) are accurate lower bounds and differ only
marginally (two overlapping lower curves). This indicates the dominance of
randomness over correlations.\hfill \hspace{1.0cm}}
\label{fig:D}
\end{figure}

To examine the effect of correlations on the apparent FD we next consider a
model in which rods are randomly located as before but with the restriction
that the rods cannot overlap. The system is assumed to be at equilibrium. The
excluded volume effect clearly creates correlations in the positions of the
rods. This example is also fully solvable\cite{me:random-model} and represents
an important class of systems with correlations such as models of hard-sphere
liquids and energy spectra with level repulsion.  We will now show that the
correlation introduced by the non-overlap restriction merely {\em modifies}
the apparent fractal character of the system.  For this case, the expected
number of intersected boxes is:\cite{me:random-model}

\begin{equation}
\langle N(r)\rangle = {1 \over r} \left(1-(1-\eta) \left( 1- {r \over
{1-\eta}} \right)^{M} \right).
\label{eq:Nhard}
\end{equation}

\begin{figure}
\psfig{figure=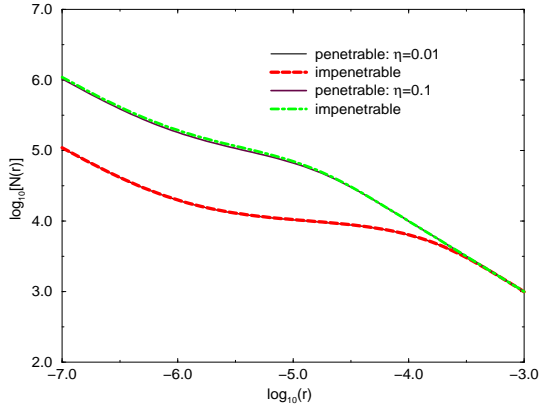,width=12cm,angle=270}
\vskip -1.6cm
\caption{
Comparison of box-counting predictions in penetrable and impenetrable rods
cases. The results for penetrable [Eq.(\protect\ref{eq:<N>})] and impenetrable
rods [Eq.(\protect\ref{eq:Nhard})] virtually coincide for $\eta \leq
10^{-2}$ (lower two curves). For $\eta=0.1$ a barely noticeable difference develops (upper two curves). In both
cases $d/L=10^{-6}$.\hfill \hspace{1.0cm}}
\label{fig:impenetrable}
\end{figure}

\noindent Fig.\ref{fig:impenetrable} shows the number of intersected boxes
$\langle N(r)\rangle$ vs. $r$ both with [Eq.(\ref{eq:<N>})] and without
[Eq.(\ref{eq:Nhard})] overlap. The behavior in the two cases is qualitatively
similar and virtually indistinguishable for low
coverages. Fig.\ref{fig:impenetrable} thus demonstrates that the apparent
fractal behavior due to randomness is only slightly modified by moderate
correlations.  As in the overlapping rods case, we can now use Eq.(\ref{eq:D})
(with the slope calculated at $r=r_e$) to calculate a lower bound for the
apparent FD. The result (for large $M$),

\begin{equation}
D = 1- { {\eta \sqrt{{1/\eta} -1}}
\over {\exp \left(\sqrt{\eta/(1-\eta)} \right) - (1-\eta) } }
\label{eq:Dhard}
\end{equation}

\noindent is shown in Fig.\ref{fig:D}. The important observation is that for a
broad range of low coverages the apparent FD's of penetrable and impenetrable
rods nearly overlap. This is the relevant range for fractal measurements and
therefore we find that correlations of the type considered here have little
effect on the apparent fractal nature of the system.

\section{Fat-Fractal Analysis}
\label{fat-fractal}

In this section we treat the penetrable spheres model for the case of
two-dimensional (2D) disks, from the point of view of fat-fractal analysis.  A
fat fractal is defined as ``A set with a fractal boundary and finite Lebesgue
measure''.\cite{Umberger:2} The fat-fractal approach is natural for our model,
since the set of disks clearly has non-zero measure. Fat-fractal analysis can
be performed on experimental data (but rarely is) in those cases where the
resolution of the measurement device is finer than the lower cut-off, which is
required for a knowledge of the measure of the studied set. An example is
helium scattering.\cite{me:fractals} In the present case we show that the
measure of the set of disks can be found analytically.  In order to measure
the fat-fractal scaling exponent $\gamma$, one performs, as in the standard
fractal analysis, a box-counting procedure:

\begin{equation}
\gamma = \lim_{r \rightarrow r^*} {{\log[A(r)]} \over \log(r)}\:;\:\:\:\: A(r)
\equiv r^2 N(r) - \mu_0,
\label{eq:gamma}
\end{equation}

\noindent where $\mu_0$ is the normalized Lebesgue measure of the set. The
fractal dimension itself is given by

\begin{equation}
D_{ff} = 2-\gamma .
\label{eq:Dff}
\end{equation}

\noindent In the nonlinear dynamical systems literature, where fat fractals
were first introduced,\cite{Umberger} $r^*=0$. In the context of real-space
sets, there exists a lower cutoff $r_0 > 0$, and hence also $r^* > 0$. One
should bear this in mind whenever fractal theory is applied to real-space
systems with an inherent non-vanishing smallest scale.

Consider then again a system of $M$ uniformly randomly positioned disks of
equal radius $R=d/2$, located at low 2D coverage $\eta_2$ given by:

\begin{figure}
\psfig{figure=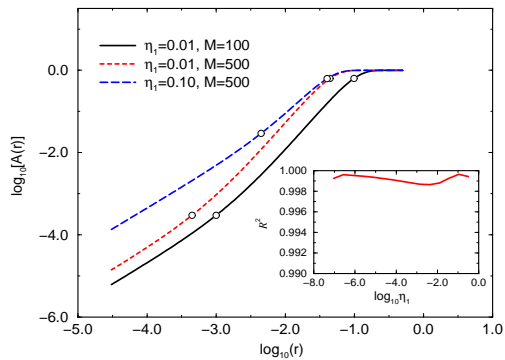,width=12cm,angle=270}
\vskip -1.6cm
\caption{
Fat fractal analysis of random disks. Analytical results of
Eq.(\protect\ref{eq:gamma}) for $A(r)$ are shown at three combinations of
coverages $\eta_1$ and disk numbers $M$. Circles indicate the positions of the
cutoffs according to Eqs.(\protect\ref{eq:r0}),(\protect\ref{eq:r1}). Inset:
Linear regression coefficient ${\cal R}^2$ for regression in-between the
cutoffs.\hfill \hspace{1.0cm}}
\label{fig:fat-fractal}
\end{figure}

\begin{equation}
\eta_2 = M \pi R^2/L^2 = (\pi/4)\eta_1^2 ,
\label{eq:eta}
\end{equation}

\noindent on a surface of area $L^2$. The effective ``1D coverage''

\begin{equation}
\eta_1=\sqrt{M}2R/L,
\label{eq:eta1}
\end{equation}

\noindent is defined for convenience of comparison with results in
1D and 3D. In order to find $\mu_0$, imagine that the surface is initially
empty, and randomly choose a point on it. Next locate a disk of radius $R$ at
a random position on the surface. The probability that it does not include the
chosen point is proportional to the free area, namely $q_1 = (L^2-\pi
R^2)/L^2$. The next disk is also positioned completely randomly, so that the
probability for the point to be outside of both disks is just
$q_1^2$. Clearly, after random placement of $M$ disks, the point will lie in
the uncovered region with probability $q_1^M$, and therefore will be in the
disk-covered region with probability $p_M = 1-(1-\pi R^2/L^2)^M$. On the other
hand, this probability is just the expectation value of the normalized disk
union area, $\mu_0/L^2$. Thus for large enough $M$:\cite{Weissberg,Torquato:3}

\begin{equation}
\mu_0 = \left[ 1-(1-\pi R^2/L^2)^M \right] L^2 .
\label{eq:mu0}
\end{equation}

\noindent A modified argument can be used to evaluate the BC function for our
basic model.\cite{me:random-model} The result for the expected number of
occupied boxes is:

\begin{equation}
N(r) = {L^2 \over r^2} \left[ 1-\left(1 - r^2 - 4r\,R/L - \pi (R/L)^2
\right)^M \right] .
\label{eq:N}
\end{equation}

\noindent Simulations\cite{me:random-model} (not shown here) confirm the 1D
version of this result to excellent accuracy for $M$ as small as 100.  Taken
together, Eqs.(\ref{eq:mu0}),(\ref{eq:N}) determine the fat-fractal exponent
$\gamma$, using Eq.(\ref{eq:gamma}). Analytical results are shown in
Fig.\ref{fig:fat-fractal} for three $\eta_1 / M$ pairs. The effect of changing
$M$ at constant coverage (solid and short-dashed lines) is a rigid translation
of the curve in the plane. This implies that the coverage is the important
parameter in determining the slope i.e., the FD. Circles indicate the
positions of the cutoffs according to
Eqs.(\protect\ref{eq:r0}),(\protect\ref{eq:r1}). Beyond the lower cutoff the
slope tends to 1, beyond the upper cutoff -- to 0. In-between the cutoffs, a
nearly straight line is observed, in agreement with apparent fractal
behavior. In order to find $\gamma$ it remains to determine the point
$r^*$. For disks, in analogy to the discussion for rods, the cutoffs are given
by:

\begin{figure}
\psfig{figure=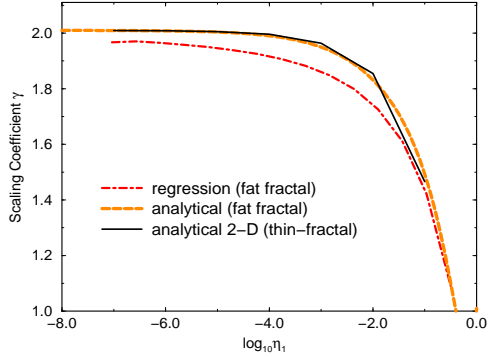,width=12cm,angle=270}
\vskip -1.8cm
\caption{
Analytical and regression slope between	the cutoffs (dashed and long dashed)
of fat fractal analysis, and $2\!-\!D$ of ``thin'' fractal
analysis.\hfill \hspace{1.0cm}}
\label{fig:fat-fractal-dims}
\end{figure}

\begin{eqnarray}
r_0 &=& 2R/L \nonumber \\
r_1 &=& 1/\sqrt{M}-2R/L .
\label{eq:r0+r1}
\end{eqnarray}

\noindent As in Sec.\ref{model} we choose $r^*$ as the estimated middle point
of the scaling range,

\begin{equation}
r^* = r_e = \sqrt{r_0\,r_1} ,
\label{eq:rm}
\end{equation}

\noindent and find $\gamma$ by evaluating the logarithmic derivative of
$A(r)$ there. The result is:

\begin{equation}
\gamma = {{d \log[A(r)]} \over {d \log(r)}}\left|_{r_e}\right. =
{{2\eta_1(1-\eta_1+\sqrt{\eta_1-\eta_1^2})} \over
{\exp[\eta_1(1-\eta_1+2\sqrt{\eta_1-\eta_1^2})]-1}} .
\label{eq:gamma-analytical}
\end{equation}

\begin{figure}
\psfig{figure=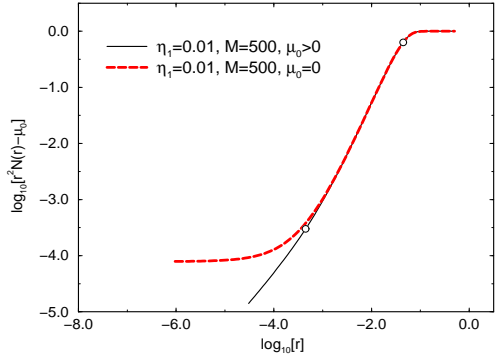,width=12cm,angle=270}
\vskip -1.8cm
\caption{
Log-log plot of $A(r) = r^n N(r) - \mu_0$ (solid line)
and just $r^n N(r)$ (dashed line). Inside the physical range, the measure
$\mu_0$ has little effect. Circles indicate the cutoffs.\hfill \hspace{1.0cm}}
\label{fig:artifact}
\end{figure}

\noindent This result is compared in Fig.\ref{fig:fat-fractal-dims} to the
result of the regular (``thin'') fractal analysis, for which we
found:\cite{me:random-model}

\begin{equation}
2 - D = 
{
{ {1 \over 2}[ \eta_1 \sqrt{2\eta_1-\eta_1^2} + 2\eta_1-\eta_1^2 ] }
\over
{ \exp\{ \eta_2 + {1 \over 2} [ \eta_1 \sqrt{2\eta_1-\eta_1^2} + {1 \over 2} (2\eta_1-\eta_1^2) ] \} -1 }
} .
\label{eq:D-thin}
\end{equation}

\noindent The two curves differ only slightly. The analytical fat-fractal
result is also compared in Fig.\ref{fig:fat-fractal-dims} to the procedure
followed in typical experimental analysis of fractal scaling data: a linear
regression between the physical cutoffs ($r_0$ and $r_1$ in our case). The
trend is similar, and the agreement is quite good for the higher coverages. In
any case the analytical Eq.(\ref{eq:gamma-analytical}) serves as an accurate
upper bound to the expected regression result for $\gamma$. The corresponding
regression coefficient ${\cal R}^2$ (inset of Fig.\ref{fig:fat-fractal}) does
not fall below 0.9985 which indicates a very high quality regression,
certainly by experimental standards. Note that ${\cal R}^2$ remains very high
even for $\eta_1 < 10^{-2}$ (i.e., a scaling range $\Delta_e > 2$). This is a
wider range than found for the 1D version of ``thin'' fractal analysis. There
the apparent fractality was observed in a range of 1-2 decades if $\eta <
10^{-1}$ and ${\cal R}^2>0.97$ are required (Fig.\ref{fig:range}).  This range
improvement is {\em not} a direct outcome of the inherent fractality we found
between the cutoffs, but is due to the differences in the response of the linear
regression procedure to Eq.(\ref{eq:gamma}), in comparison to Eq.(\ref{eq:D}): It is
the multiplication of the BC function by $r^2$ in the former which is
responsible for the improved range effect, through the increase in the slope of the log-log
plot. One should be cautious therefore to eliminate slope-biases in analyses
of scaling properties on a log-log plots. Furthermore, as clearly seen in
Fig.\ref{fig:artifact}, the essence of the fat fractal analysis, namely the
subtraction of the measure $\mu_0$ of the set, has essentially no effect on
the slope and on the location of the cutoffs and thus does not provide us in
this case with added information. Being left then with the choice between
$N(r)$ and $r^2 N(r)$, there does not seem to be a clear reason to opt for the
latter. We conclude that for low density systems, such as in this report, fat
fractal analysis is not necessary.

\section{Conclusions}
In summary, we have shown that random structures, which are generic in
experimental situations where only the first moment of a distribution is
determined, give rise to apparent fractal behavior within physically relevant
cutoffs, with a non-universal FD. Although this is not a mathematically
rigorous fractality, in the sense that the scaling is not strictly a power
law, it is a {\em physical} fractality: It satisfies the conditions of
high-quality linear regression in the physically relevant range of
observation. Since experiments rarely observe a perfect power law, we believe
that the possibility of {\em approximate} scaling should be considered in
theoretical models, if a more complete understanding of the experimental
fractal data is to be achieved. It is likely that some of this data does in
fact not reflect the existence of an exact power law, but rather an
approximate power law between cutoffs with a weak inflexion point in the
log-log plot. The present model and its approximate scaling properties hint
that this may be the case, e.g., for porous media. Moderate correlations have
little effect on the apparent fractal properties and even in their presence it
is still the underlying randomness that is the main contributor to the
apparent power-law scaling relation. Elsewhere we showed that these results
remain practically unchanged for higher dimensions and for a variety of size
distribution profiles of the elementary building blocks.\cite{me:random-model}
We thus propose to consider randomness as a possible common source for
apparent fractality.

\section*{Acknowledgments}
We would like to thank R.B. Gerber, D. Mukamel and G. Shinar for very helpful
discussions. D.A. is a member of the Fritz Haber Research Center for Molecular
Dynamics and of the Farkas Center for Light Energy Conversion.


\begin{thebibliography}{10}
\bibitem{higherorder}
{We have also applied the ``Minkowski Sausage'' technique. In one dimension the
  two methods naturally provide identical results while we found that in two
  and three dimensions they slightly differ.\cite{me:random-model} Both methods
  measure $D_0$. Elsewhere we will treat the case of $D_q$, $q \neq 0$.}
\bibitem{sketch}
{The idea is to pick one box and then randomly deposit $M$ rods of size $d$ on
  the unit interval. The probability that the chosen box remains empty is $q =
  [ 1 - (r+d/L) ]^{M}$ and thus the probability that it will intersect at least
  one rod is $p=1-q$.}
\bibitem{goodfit}
{This excellent agreement appears for all scales in $r$ including the ranges of
  slope=1. This was also confirmed for other choices of $M$ and $d$ values and
  even for very large $\eta$ where apparent fractal behavior is not seen.}
\end{thebibliography}
\end{document}